\documentclass{iau}
\usepackage{graphicx}

\usepackage[numbers]{natbib}
\title[A Possible Dynamical History for the Fomalhaut System] 
{A Possible Dynamical History for the Fomalhaut System}

\author[V. Faramaz]   
{Virginie Faramaz$^1$
}

\affiliation{$^1$Instituto de Astrof\'isica - Pontificia Universidad Cat\'olica de Chile\\email: {\tt vfaramaz@astro.puc.cl}
}

\pubyear{2015}
\volume{314}  
\pagerange{119--126}
\jname{Young Stars \& Planets Near the Sun}
\editors{J. H. Kastner, B. Stelzer, \& S. A. Metchev, eds.}
\begin{document}

\maketitle

\begin{abstract}
Fomalhaut b was long thought to shape the eccentric debris belt in the Fomalhaut system, but its orbit was found to be too eccentric for it to be the dominant belt-shaping perturber. This indicates that Fomalhaut b is Earth-sized at most and that the belt-shaping perturber, hereafter named Fomalhaut c, remains to be discovered. In addition, since its orbit more or less crosses that of Fomalhaut b, it also indicates that the current configuration of the system is transient and was reached recently. In this talk, we show that this current configuration can be explained if Fomalhaut c is Saturn- to Neptune-sized, and Fomalhaut b originates from a mean-motion resonance with Fomalhaut c. 
\end{abstract}

\firstsection 
\section{Introduction}

\begin{figure}[b]
\begin{center}
 \includegraphics[width=0.4\textwidth,clip]{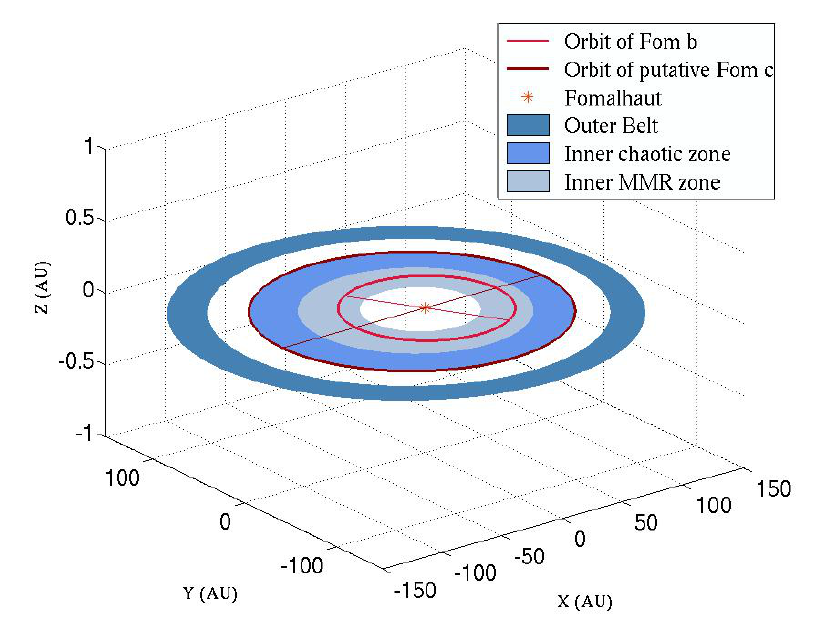}     
 \includegraphics[width=0.4\textwidth,clip]{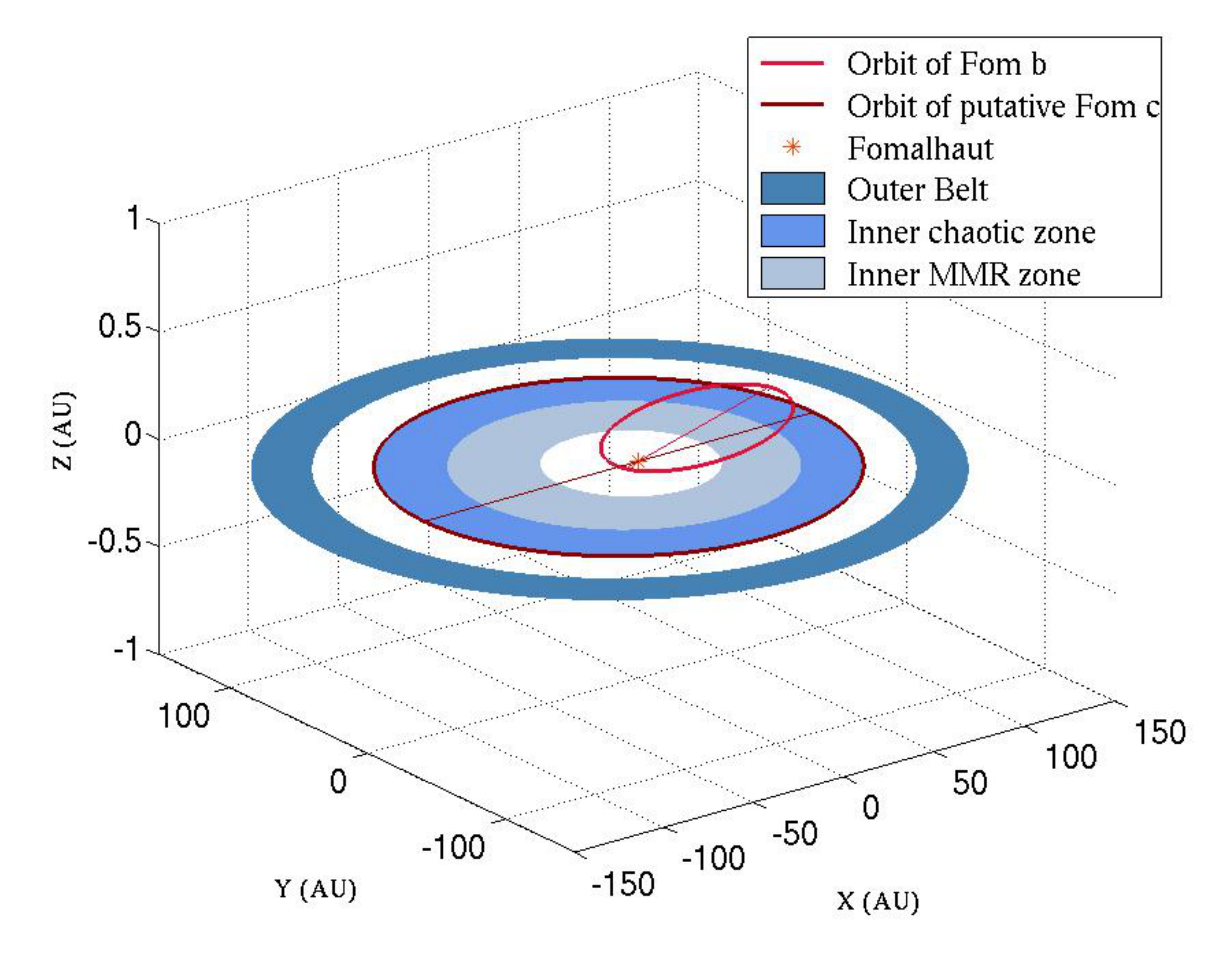} 
 
   \includegraphics[width=0.4\textwidth,clip]{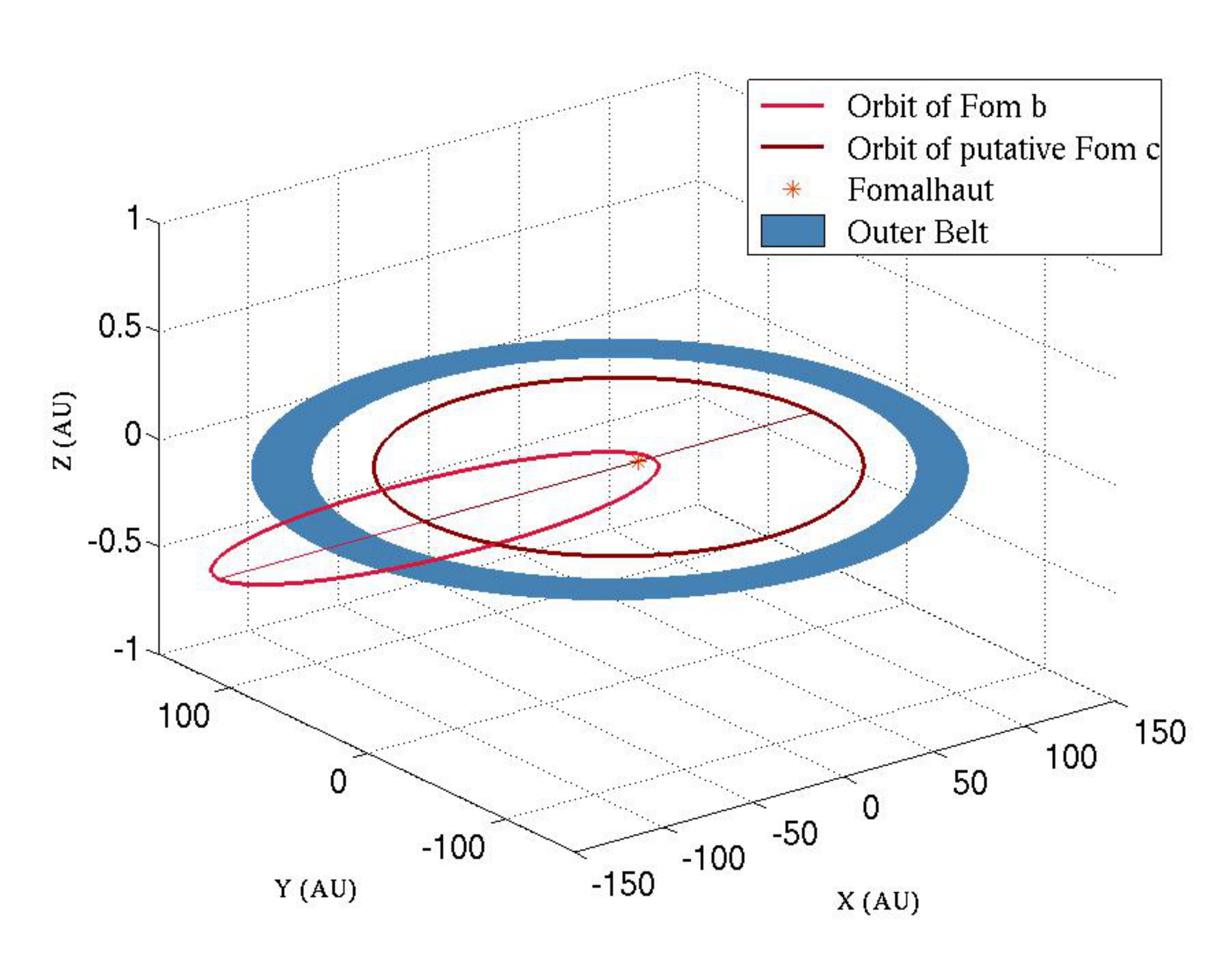} 
 \caption{{\bf Top-left:} Probable initial configuration of the Fomalhaut system. Fom b is in MMR with the belt-shaping eccentric Fom c. {\bf Top-right:} Probable intermediate configuration of the Fomalhaut system. MMRs with an eccentric perturber generate very eccentric orbits, which leads Fom b to cross the chaotic zone of Fom c and be scattered on its current orbit. {\bf Bottom:}  Probable current configuration of the Fomalhaut system.}
  \label{faramaz:fig2}
\end{center}
\end{figure}

Fomalhaut ($\alpha\:$Psa) is a 440 Myr old A3V star, located at 7.7\,pc \citep{2007ASSL..350.....V,2012ApJ...754L..20M}.  Fomalhaut is surrounded by an eccentric dust ring ($\mathrm{e}=0.11\pm 0.01$) \citep{2005Natur.435.1067K}. This eccentric shape hinted at the presence of a massive body orbiting inside the belt on an eccentric orbit, dynamically shaping the belt \citep{2006MNRAS.372L..14Q,2005ApJ...625..398D}. This hypothesis was apparently confirmed by the direct detection of a companion near the inner edge of the belt, Fomalhaut~b \citep[hereafter Fom b;][]{2008Sci...322.1345K}. 
However, orbital fitting for this perturber has revealed a highly eccentric orbit that crosses the belt. In addition, it is close to apsidal alignement with the belt. Since such an eccentric orbit would create a significant apsidal misalignement, Fom b cannot be responsible for the disk shaping \citep{2013ApJ...775...56K,2014A&A...561A..43B,2015MNRAS.448.3679P}.
The most straightforward solution to this apparent paradox is to suppose the presence of a yet undetected body in the system, hereafter named Fom c, which would be much more massive than Fom b. Consequently, Fom c would be dynamically predominant and responsible for the belt shaping. This is supported by recent dynamical or photometric studies which suggest that Fom b is no more than Earth- or Super-Earth sized \citep{2014A&A...561A..43B,2012ApJ...747..116J,2013ApJ...769...42G}. However, in this configuration, which is illustrated in the bottom panel of Fig.~\ref{faramaz:fig2}, the orbit of the putative belt-shaping planet Fom c would be crossed by that of Fom b. Such a two-planet system is highly unstable, and would indicate that Fom b must have been perturbed recently, potentially by Fom c \citep{2014A&A...561A..43B}.

\section{Investigating the dynamics of a two-planet system}

Investigation of the dynamics of this two-planet system in \citet{2015A&A...573A..87F}, where Fom c is a massive belt-shaping body and Fom b a much less massive body originating from the inner parts of the system, has revealed a possible three-step dynamical scenario which can explain both why Fomalhaut b is on such an eccentric orbit and why it was set on it recently:

\begin{enumerate}
\item  \emph{Mean-Motion Resonances between Fom b and the suspected Fom c :} Fom b is likely to have formerly resided in an inner mean-motion resonance (MMR) with the additional planet, as illustrated in the top-left panel of Fig.~\ref{faramaz:fig2}. MMRs with an eccentric perturber such as the belt-shaping Fom c induce a gradual eccentricity increase, which can cause Fom b to cross the chaotic zone of Fom c, where it can then be scattered by Fom c on its current orbit (top-right panel of Fig.~\ref{faramaz:fig2}). The dynamical timescale involved in this process, that is, the typical time necessary for Fom b to reach a sufficient orbital eccentricity from its MMR position and be scattered on its current orbit, strongly depends on the mass of the putative Fom c. In particular, the scattering event can be delayed on timescales comparable to the age of the system with a Neptune- or Saturn-sized Fom c, which would explain why Fom b was recently set on its orbit.
\item \emph{Close encounter with the suspected Fom c:} inspection of the close encounters between Fom b and Fom c reveals that these can set Fom b on an orbit with semi-major axis compatible with that of Fom b, but that they also preferentially produce orbits which are not eccentric enough to be compatible with that of the observed one \citep[$a = 81-415\,$AU and $e = 0.69-0.98$, at the 95\% confidence level][]{2014A&A...561A..43B}.
\item \emph{Secular evolution with the suspected Fom c:} an additional eccentricity increase can be provided when Fom b is under the secular influence of the eccentric Fom c, which is indeed mainly expected at semi-major axes with $a = 81-415\,$AU. However, this eccentricity increase is accompanied by an apsidal alignement with the belt-shaping Fom c, and thus with the belt, which may explain the tendency for the observed orbit to be apsidally aligned with the belt.
\end{enumerate} 

\begin{figure}[b]
\begin{center}
 \includegraphics[width=0.85\textwidth,clip]{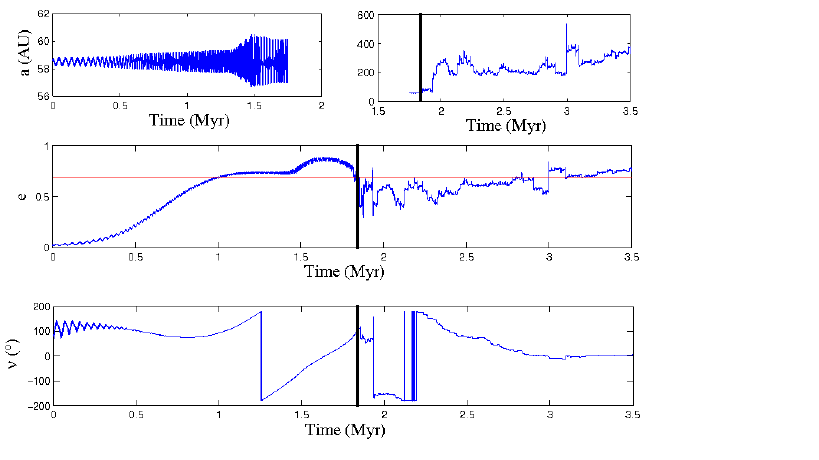} 
 \caption{{\bf Example of the three-step process that may have placed Fom b on its current orbit.} 
  We display the evolution in time of the semi-major axis $a$, eccentricity $e$, and longitude of periastron $\nu$ of a massless test particle initially in 5:2 MMR with a 3$\mathrm{M_{Jup}}$ Fom c, with semi-major axis 108.6 AU and orbital eccentricity 0.1. Note that this process can be generated via several other MMRs. The test particle endures a three-step dynamical evolution, starting with resonant evolution, where its semi-major axis suffers small oscillations around the exact resonant location, and its eccentricity largely increases, while co-evolving with the longitude of periastron. The vertical black line at $\sim 2\,$Myr indicates the second step of the process, that is, a close encounter with Fom c when the highly eccentric orbit of the test particle crosses the chaotic zone of Fom c. Note that this delay of several Myr with a Jupiter-sized Fom c increases up to several 100 Myr with a Saturn- or Neptune-sized Fom c. The semi-major axis of the test particle is comparable with this of Fom b after the close encounter, but its eccentricity remains smaller than 0.69 \emph{(horizontal red line)}, and thus is incomparable with that of Fom b. The third step consists mainly in a secular evolution of the test particle with the eccentric Fom c, although its orbit endures small chaotic variations. This secular evolution allows the eccentricity to increase and become greater than 0.69, which occurs when there is an apsidal alignement between the perturber and the test particle, that is, when the longitude of periastron of the test particle is close to zero.}
  \label{faramaz:fig3}
\end{center}
\end{figure}

The whole process is summarized and illustrated in Fig.~\ref{faramaz:fig3}.

In addition, when an eccentric planet such as Fom c coexists with km-sized planetesimals, orbits such as that of Fom b can be expected to be generated with great efficiency via the three-step mechanism detailed above \citep{2015A&A...573A..87F}. Therefore, one should probably expect the Fomalhaut system to contain a broad population of solid bodies on highly eccentric orbits that closely approach the star at periastron. These bodies would feed the inner parts of the system with dust, creating hot or warm inner belts. This is extremely interesting in the context of the Fomalhaut system, since both a warm and a hot inner belt were detected  \citep{2013A&A...555A.146L}. 

\section{Conclusions}

The study of the Fomalhaut system has revealed a robust process by which orbits such as that of Fom b naturally result from interactions between an eccentric massive perturber such as the suggested Fom c -- assumed to shape the outer belt of this system -- and a much less massive body, such as a smaller planet or a planetesimal. In addition, this process involves a delay in the production of Fom b-like orbits, which can be greater than 100 Myr if the eccentric massive perturber is Neptune- to Saturn-sized. This can provide an explanation both for the shape of the outer belt and the current dynamical status of Fom b. In addition, it may be at the origin of inner belts in the Fomalhaut system, and provide a solution to the presence of unusual high levels of dust in the vicinity of a significant number of stars with age $> 100\,$Myr (Faramaz et al., in prep).

\bibliographystyle{aa}  
\bibliography{biblio}

\begin{discussion}

\discuss{E. Mamajek}{Fomalhaut A is part of a triple star system. Did you consider one of the stellar companions could be responsible for the shape of the debris disk instead of an unseen Fom c?}

\discuss{V. Faramaz}{Indeed, \citet{2014MNRAS.442..142S} argued that Fom C could excite the eccentricities of the disk at the observed value. However increasing eccentricities is not sufficient to create a well-defined eccentric ring, which is an extended structure. Providing first that Fom C is coplanar to the debris disk of Fom A, which is unknown, the process takes several precession timescales. With a separation of $\sim 200\,$kAU between Fom A and Fom C \citep{2013AJ....146..154M}, this precession timescale is of the order of Gyr, that is, much longer than the age of the Fomalhaut system. This is why the possibility that Fom C is responsible for the shape of the debris disk of Fom A is discarded by \citet{2015A&A...573A..87F}.}

\discuss{Unidentified conference participant}{What do you think about the scenario by \citet{2015MNRAS.448..376N} in which Fom b is actually a background neutron star?}

\discuss{V. Faramaz}{It is not a possibility to exclude completely, however, this scenario fails to explain the fact that the orbit of Fom b was found to be nearly coplanar with the debris ring, as found first by \citet{2013ApJ...775...56K}, and confirmed since by \citet{2014A&A...561A..43B} and \citet{2015MNRAS.448.3679P}. The neutron star scenario was also motivated by the fact that the lack of giant planets in the inner parts of the system ruled out a scenario in which Fom b was scattered on its current orbit.  However, as shown by \citet{2015A&A...573A..87F}, there is no need for such giant planets for Fom b to be scattered, and the inferred belt-shaping planet itself, which can be as small as a Neptune, can explain the current configuration of the system.}

\end{discussion}

\end{document}